\documentclass[12pt]{iopart}

\usepackage{graphicx}

\def\lesssim{\mathrel{\rlap{\lower4pt\hbox{\hskip1pt$\sim$}}}<}
\def\gtrsim{\mathrel{\rlap{\lower4pt\hbox{\hskip1pt$\sim$}}}>}

\begin{document}

\title[Dark matter and the first stars] {Dark Stars: A New Study of  the First Stars in the Universe}

\author{Katherine Freese} 

\address{Michigan Center for Theoretical Physics, University of Michigan, Ann Arbor, MI 48109}
\ead{ktfreese@umich.edu}

\author{Peter Bodenheimer} 

\address{Astronomy Dept., University of California, Santa Cruz, CA 95064}
\ead{peter@ucolick.org}

\author{Paolo Gondolo} 

\address{Physics Dept., University of Utah, Salt Lake City, UT 84112}
\ead{paolo@physics.utah.edu}

\author{Douglas Spolyar}
 
\address{Physics Dept., University of California, Santa Cruz, CA 95064}
\ead{dspolyar@physics.ucsc.edu}


\begin{abstract}
We have proposed that the first phase of stellar evolution in the history of the Universe may be Dark Stars (DS),  powered by dark matter  heating rather than by nuclear fusion.
Weakly Interacting Massive Particles, which may be their own
antipartners, collect inside the first stars and annihilate to produce
a heat source that can power the stars.  A new stellar phase results,
a Dark Star, powered by dark matter annihilation as long as there is
dark matter fuel, with lifetimes from millions to billions of years.  We find that
the first  stars are very bright ($\sim 10^6 L_\odot$) and cool ($T_{surf} < 10,000$K) during the DS phase,
and grow to be very massive (500-1000 times as massive as the Sun).  These results differ markedly from the standard
 picture in the absence of DM heating, in which the maximum mass is about  140$M_\odot$ and the temperatures are much hotter ($T_{surf} > 50,000$K);
 hence DS should be observationally distinct from standard Pop III stars.
 Once the dark matter fuel is exhausted, the DS becomes a heavy main sequence star;
these stars eventually collapse to form massive black holes that
 may provide seeds for  supermassive black holes observed at
 early times as well as explanations for recent ARCADE data  and for intermediate black holes.  

\end{abstract}

\maketitle

\section{Introduction}

We have proposed \cite{sfg} (hereafter Paper I) a new phase
of stellar evolution: the first stars to form in the Universe may be Dark Stars, powered
by dark matter heating rather than by fusion.  Here dark matter, while
constituting a negligible fraction of the star's mass, provides the energy source
that powers the star.  The first stars in the Universe  
 mark the end of the cosmic dark ages, provide the enriched gas required for later
stellar generations, contribute to reionization, and may be precursors to
black holes that coalesce and power bright early quasars.  One of the outstanding
problems in astrophysics is to investigate the mass and properties of these first stars. Our results differ in important ways from the standard picture of first stars without DM heating.

Weakly Interacting Massive Particles (WIMPs)
 are the best motivated dark matter candidates.  WIMP annihilation in the
early universe provides the right abundance today to explain the dark
matter content of our universe. This same annihilation process will
take place at later epochs in the universe wherever the dark matter
density is sufficiently high to provide rapid annihilation.  The first
stars to form in the universe are a natural place to look for
significant amounts of dark matter annihilation, because they form at
the right place and the right time. They form at high redshifts, when
the universe was still substantially denser than it is today, and at
the high density centers of dark matter haloes. 
 
The first stars form inside dark matter (DM) haloes of $10^6 M_\odot$
(for reviews see e.g. \cite{Ripamonti:2005ri, Barkana:2000fd, Bromm:2003vv, 
Yoshida:2008gn}; see also \cite{ABN, Yoshida06}.)
One star is thought to form inside one such DM halo. The first stars may
play an important role in reionization, in seeding supermassive black
holes, and in beginning the process of production of heavy elements in
later generations of stars.  

It was our idea to ask, what is the
effect of the DM on these first stars?  We studied the behavior of
WIMPs in the first stars, and  found that they 
can radically alter the stellar evolution.  The annihilation products of the
dark matter inside the star can be trapped and
deposit enough energy to heat the star and prevent it from further
collapse.  A new stellar phase results, a Dark Star, powered
by DM annihilation as long as there is DM fuel, for millions to billions of years.


\subsection{Weakly Interacting Dark Matter}

\label{sec:WIMPs}

WIMPs are  natural dark matter candidates from particle physics.
These particles, if present in thermal abundances in the early
universe, annihilate with one another so that a predictable number of
them remain today.  The relic density of these particles is
\begin{equation}
\Omega_\chi h^2 = (3 \times 10^{-26} {\rm cm}^3/{\rm sec})
/ \langle \sigma v \rangle_{ann}
\end{equation}
where the annihilation cross section $\langle \sigma v \rangle_{ann} $
of weak interaction strength automatically gives the right answer, near the
WMAP  \cite{Komatsu:2008hk} value $\sim 23\%$.
This coincidence is known as "the WIMP miracle" and is the reason why
WIMPs are taken so seriously as DM candidates.  The best WIMP
candidate is motivated by Supersymmetry (SUSY): the lightest
neutralino in the Minimal Supersymmetric Standard Model
(see the reviews by \cite{jkg, jkgb, jkgc, Bertone_etal2004}).

This same annihilation process is also the basis for
DM indirect detection searches.  The first paper discussing annihilation in stars was \cite{Krauss:1985ks}; the first papers suggesting searches for annihilation products of WIMPs in the Sun
were by Silk {\it et al} \cite{SOS}; and in the Earth
by Freese \cite{freese} as well as Krauss, Srednicki and
Wilczek \cite{ksw}.  Other studies of WIMPs in today's stars 
(less powerful than in the first stars) include \cite{bouquet, salati, moskalenko, scott1, scott2, bertone}.
This article reviews the study of WIMP annihilation as a heat source for the first stars.

As our canonical parameter values, we take $m_\chi =
100$GeV for the WIMP mass and $\langle \sigma v \rangle_{ann} = 3
\times 10^{-26} {\rm cm^3/sec}$ for the annihilation cross section
but consider a variety of masses and cross sections.

\section{Three Criteria for Dark Matter Heating}

 WIMP annihilation produces energy at a rate per
unit volume 
\begin{equation}
\label{eq:heat}
  Q_{\rm ann} = \langle \sigma v \rangle_{ann} \rho_\chi^2/m_\chi
  \linebreak \simeq  10^{-29} {{\rm erg} \over {\rm cm^3/s}} \,\,\, {\langle
    \sigma v \rangle \over (3 \times 10^{-26} {\rm cm^3/s})} \left({n_h \over {\rm
        cm^{-3}}}\right)^{1.6} \left({100 {\rm GeV}\over m_\chi}\right) 
\end{equation}
where $\rho_\chi$ is the DM energy density inside the star and $n_h$ is
the stellar hydrogen density.  Paper I \cite{sfg} outlined the three key ingredients
for Dark Stars: 1) high dark matter densities, 2) the annihilation
products get stuck inside the star, and 3) DM heating wins over other
cooling or heating mechanisms.  These same ingredients are required
throughout the evolution of the dark stars, whether during the
protostellar phase or during the main sequence phase.

{\bf First criterion: High Dark Matter density inside the star.}  One can see
from Eq.(\ref{eq:heat}) that the DM annihilation  rate scales as WIMP density squared,
because two WIMPs must find each other to annihilate.  Thus the annihilation is
significant wherever the density is high enough.  Dark
matter annihilation is a powerful energy source in these first stars (and not in today's stars)
because the dark matter density is high.  First, DM densities in the early universe were
higher by $(1+z)^3$.   Second, the first stars form exactly in the centers of DM haloes
where the densities are high (as opposed to today's stars which are scattered throughout
the disk of the galaxy rather than at the Galactic Center).  We assume for
our standard case that the DM density inside the $10^6 M_\odot$ DM halo initially has an 
NFW (Navarro, Frenk \& White \cite{NFW})
 profile for both DM and gas, with
substantial DM in the center of the halo (we note that we obtain qualitatively the same
result for cored haloes). Third, a further DM enhancement
takes place in the center of the halo: as the protostar forms, it deepens the potential
well at the center and pulls in more DM as well.  We have computed this
enhancement in several ways \cite{sfg}  as discussed
in the next paragraph.  Fourth, the original DS is only $\sim 1 M_\odot$; then it accretes
more baryons as well as more DM up to almost 1000 $M_\odot$, in the process increasing
the DM density inside the star.
Fifth, at later stages, we also consider possible further enhancement
due to capture of DM into the star (discussed below).

{\it Enhanced DM density due to adiabatic contraction:}
Paper I recognized a key effect that increases the DM density: adiabatic contraction (AC).  
As the gas falls into the star, the DM is gravitationally pulled along with it.   Given the initial NFW 
profile, we follow its response to the changing baryonic gravitational potential as the
 gas condenses.  Paper I used a simple Blumenthal method \cite{Blumenthal}, which assumes circular
 particle orbits  to obtain estimates of the density. Our original DM profile matched that
 obtained numerically in \cite{ABN} with $\rho_\chi \propto r^{-1.9}$, for both their earliest
 and latest profiles; see also \cite{Natarajan:2008db} for a discussion. Subsequently
 we performed an exact calculation \cite{Freese:2008hb}
using the Young method \cite{Young} which includes radial orbits, 
 and confirmed our original results (within a factor of two).  Thus we feel confident that
 we may use the simple Blumenthal method in our work.  We found
\begin{equation}
\label{eq:AC}
\rho_\chi \sim 5 {\rm (GeV/cm^3)} (n_h/{\rm cm}^{-3})^{0.81} ,
\end{equation}
where $n_h$ is the gas density. For example, due to this contraction,
at a hydrogen density of $10^{13}$/cm$^3$, the DM density is $10^{11}$ GeV/cm$^3$.
Without adiabatic contraction, DM heating in the first stars would be so small as to be irrelevant.

{\bf Second Criterion: Dark Matter Annihilation Products get stuck
  inside the star}.  In the early stages of Pop III star formation,
when the gas density is low, most of the annihilation energy is
radiated away \cite{Ripamonti:2006gr}.  However, as the gas collapses
and its density increases, a substantial fraction $f_Q$ of the
annihilation energy is deposited into the gas, heating it up at a rate
$f_Q Q_{\rm ann}$ per unit volume.  While neutrinos escape from the
cloud without depositing an appreciable amount of energy, electrons
and photons can transmit energy to the core.  We have computed
estimates of this fraction $f_Q$ as the core becomes more dense. Once
$n\sim 10^{11} {\rm cm}^{-3}$ (for 100 GeV WIMPs), e$^-$ and photons
are trapped and we can take $f_Q \sim 2/3$.

{\bf Third Criterion: DM Heating is the dominant heating/cooling
  mechanism in the star}.  We find that, for WIMP mass $m_\chi =
100$GeV (1 GeV), a crucial transition takes place when the gas density
reaches $n> 10^{13} {\rm cm}^{-3}$ ($n>10^9 {\rm cm}^{-3}$).  Above
this density, DM heating dominates over all relevant cooling
mechanisms, the most important being H$_2$ cooling \cite{Hollenbach}.

Figure 1 shows evolutionary tracks of the protostar in the
temperature-density phase plane with DM heating included
(Yoshida et al. \cite{Yoshida_etal08}), for two DM particle
masses (10 GeV and 100 GeV).  Moving to the right on this plot is
equivalent to moving forward in time.  Once the black dots are
reached, DM heating dominates over cooling inside the star, and the
Dark Star phase begins.  The protostellar core is prevented from
cooling and collapsing further.  The size of the core at this point is
$\sim 17$ A.U. and its mass is $\sim 0.6 M_\odot$ for 100 GeV mass
WIMPs.  A new type of object is created, a Dark Star supported by DM
annihilation rather than fusion.

\begin{figure}
  \includegraphics[height=.3\textheight]{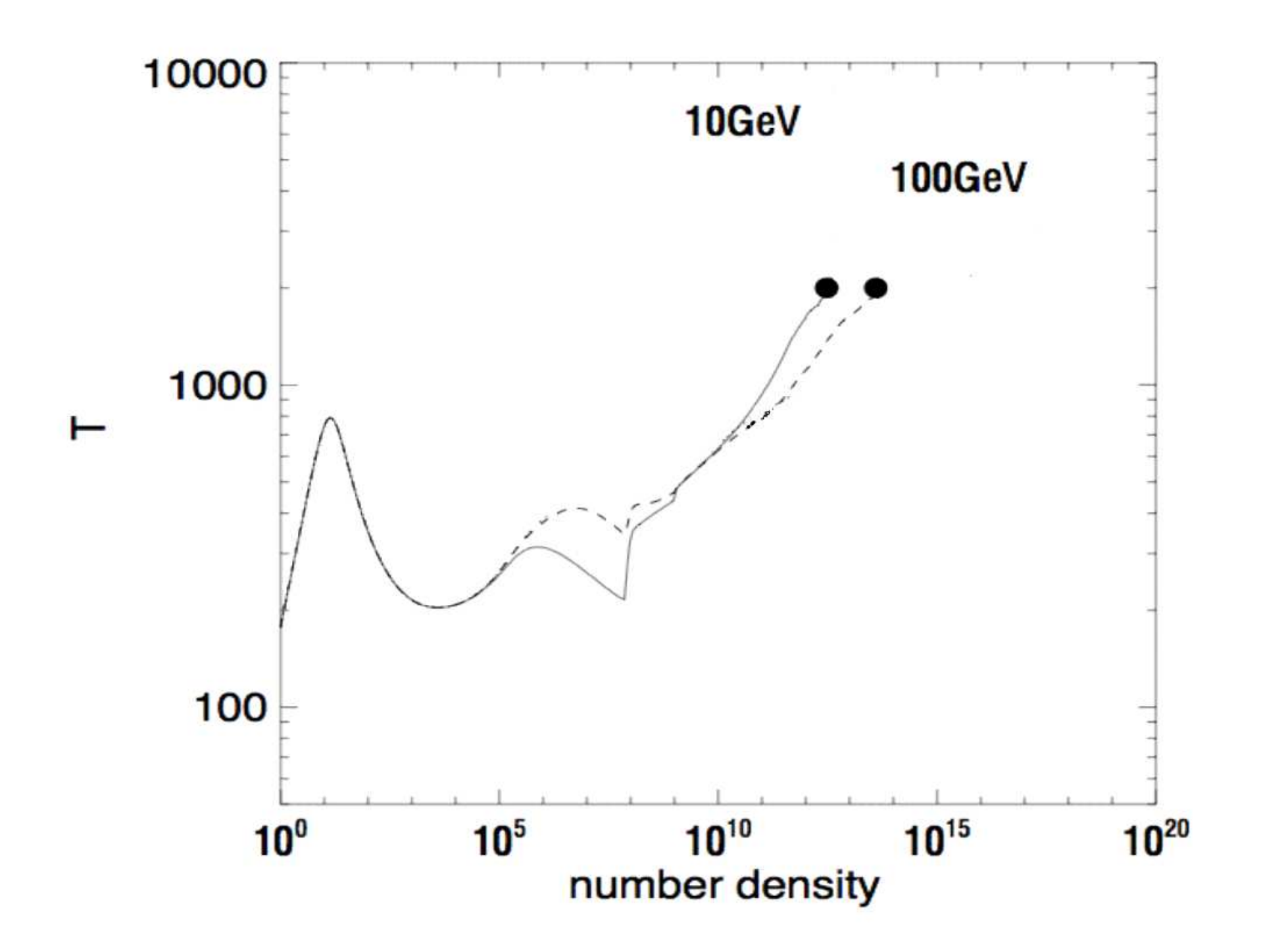}
\caption{ Temperature (in degrees K) as a function of hydrogen density
  (in cm$^{-3}$) for the first protostars, with DM annihilation
  included, for two different DM particle masses (10 GeV and 100 GeV).
  Moving to the right in the figure corresponds to moving forward in
  time.  Once the ``dots'' are reached, DM annihilation wins over H$_2$
  cooling, and a Dark Star is created.}
\end{figure}

\section{Building up the Mass}

This point is the beginning of the life of the dark star, a DM powered star
which lasts until the DM fuel runs out. 
We have found the stellar structure of the dark stars
(hereafter DS) \cite{Freese:2008wh}.  After forming with the properties
described in the previous paragraph, the DS accrete mass from the
surrounding medium.  In our paper we build up the DS mass as it grows
from $\sim 1 M_\odot$ to $\sim 1000 M_\odot$.
As further gas accretes onto the DS, more DM is pulled along with it into the star. 
At each step in the accretion process, we compute the resultant DM profile
in the dark star by using the Blumenthal 
prescription for adiabatic contraction.The DM density profile is calculated at each iteration 
of the stellar structure, so that the DM luminosity can be determined.

We allow surrounding matter from the original baryonic core
 to accrete onto the DS,with three different assumptions for the mass 
accretion: (i) $3\times 10^{-3} M_\odot/{\rm yr}$,
(ii) the variable rate from Tan \& McKee \cite{mckee} and (iii)
the variable rate from  O'Shea \& Norman \cite{Oshea}. The Tan/McKee rate decreases 
from $1.5 \times 10^{-2}$ M$_\odot/{\rm yr}$ at  a DS mass of 3 M$_\odot$ to 
$1.5 \times 10^{-3}$ M$_\odot/{\rm yr}$ at   1000 M$_\odot$. The O'Shea/Norman
rate decreases from $3 \times 10^{-2}$ M$_\odot/{\rm yr}$ at  a DS mass of 3 M$_\odot$
 to $3.3 \times 10^{-4}$ M$_\odot/{\rm yr}$ at   1000 M$_\odot$.
 As the mass increases, the DS radius adjusts
until the DM heating matches its radiated luminosity.  We find
polytropic solutions for dark stars in hydrostatic and thermal
equilibrium. We build up the DS by accreting $1 M_\odot$ at a time, always
finding equilibrium solutions.  We find that initially the DS are in
convective equilibrium; from $(100-400) M_\odot$ there is a transition
to radiative; and heavier DS are radiative.  As the DS grows, it pulls
in more DM, which then annihilates.  We continue this process until
the DM fuel runs out at $M_{DS} \sim 800 M_\odot$ (for 100 GeV WIMPs).

We have performed a complete  study of building up the dark
star mass and finding the stellar structure at each step in mass accretion.  In addition
to the heating due to DM annihilation, we included additional heat sources due to
gravitational potential energy and fusion in the later
 stages of accretion, as the DM begins to run out and the star  contracts and heats up.

Thus the energy supply for the star changes with time and comes from four major
sources: 
\begin{equation}
L_{\rm tot}=L_{DM} + L_{\rm grav} + L_{\rm nuc} + L_{\rm cap} ,
\end{equation}
 where the ingredients are the DM luminosity $L_{DM}$; gravitational contraction $L_{\rm grav}$
 (as the DM begins to run out); fusion luminosity $L_{\rm nuc}$ (once the star
 has contracted enough to reach high temperatures for fusion); and the contribution $L_{\rm cap}$
 to the luminosity due to captured DM (discussed below).  The general
thermal equilibrium condition is then that the stellar luminosity $L_\ast$ match the heat supply,
\begin{equation}
L_\ast = L_{\rm tot} . 
\end{equation}
FIgure 2 shows the different contributions to the luminosity as a function of time for the 100 GeV
case using the Tan/McKee accretion rate.  We include feedback mechanisms which can prevent further accretion. Once the stellar
 surface becomes hot enough,  when the DM is running out, the radiation can prevent
 accretion. 

Figure 3 shows the stellar structure for the case of constant accretion rate and assuming
a convective star (n=1.5); more accurate results will be found in our upcoming paper.
One can see ``the power of
darkness:'' although the DM constitutes a tiny fraction ($<10^{-3}$)
of the mass of the DS, it can power the star. The reason is that WIMP
annihilation is a very efficient power source: 2/3 of the initial
energy of the WIMPs is converted into useful energy for the star,
whereas only 1\% of baryonic rest mass energy is useful to a star via
fusion.

\begin{figure}[t]
\label{luminosity}
 \includegraphics[height=.3\textheight]{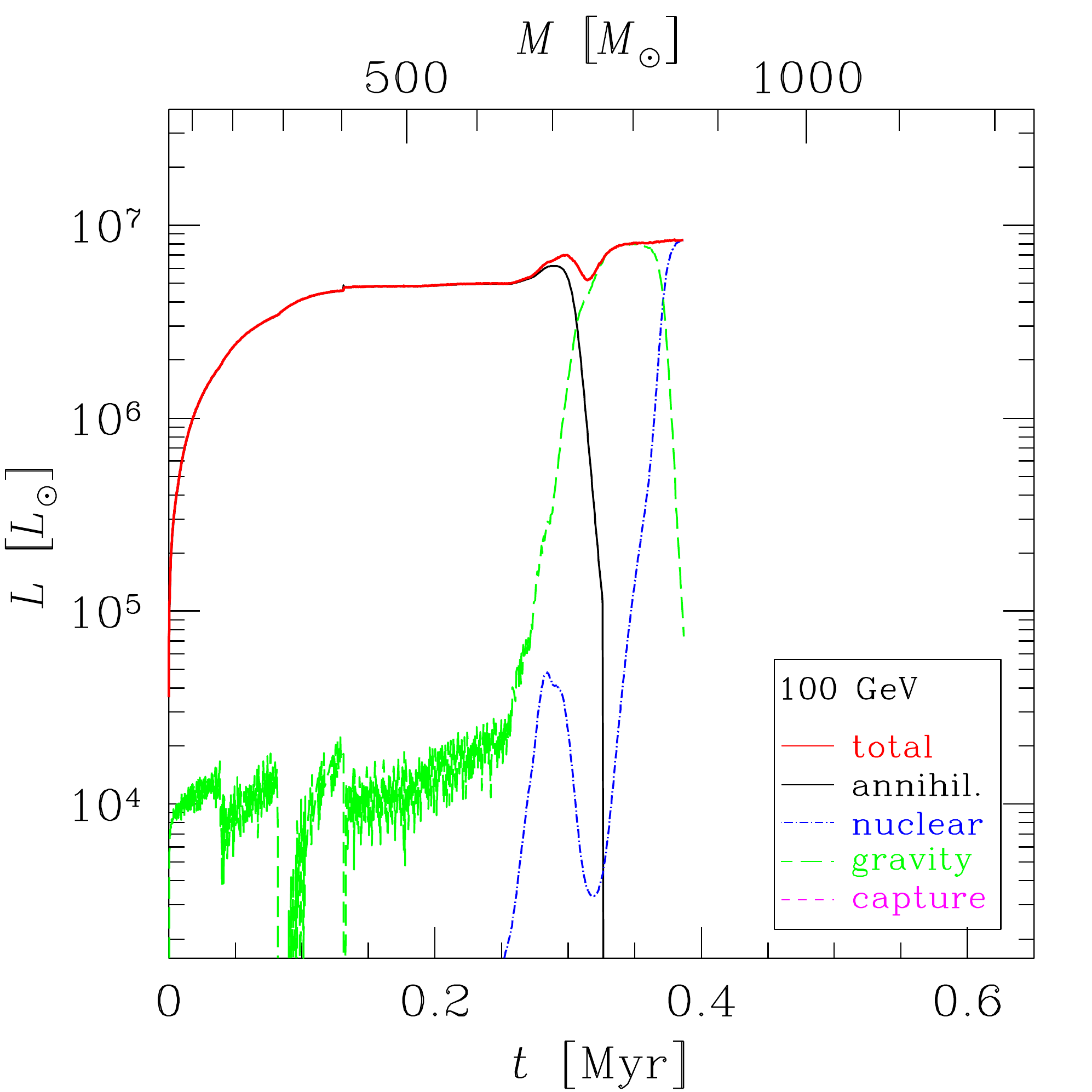}
  \includegraphics[height=.3\textheight]{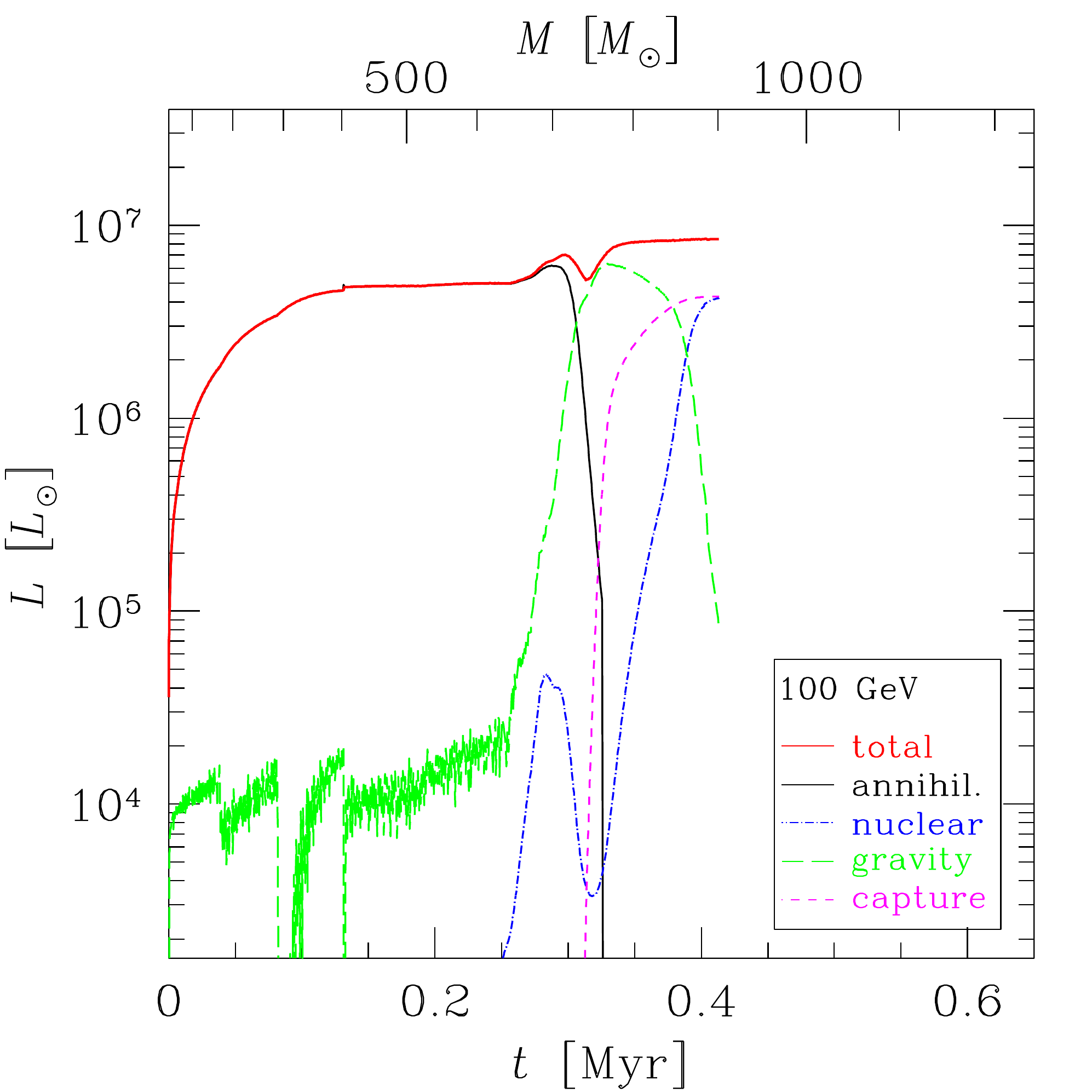}
\caption{Luminosity evolution for the 100 GeV case as a function of time
({\it lower scale}) and stellar mass ({\it upper scale}). The solid (red) 
top curve is the total luminosity.  The lower curves give the partial 
contributions of  different sources of 
energy powering the star  a) ({\it upper frame})
 without capture, and  b) ({\it lower frame}) with 'minimal' capture.  In both frames,
 the total luminosity is initially dominated by DM annihilation (the total and annihilation
 curves are indistinguishable until about 0.3 Myr after
 the beginning of the simulation); then  gravity dominates, followed by nuclear fusion.
 In the lower frame, capture becomes important at late times. 
}
\label{fig:f2}
\end{figure}

\begin{figure}
 \includegraphics[height=.3\textheight]{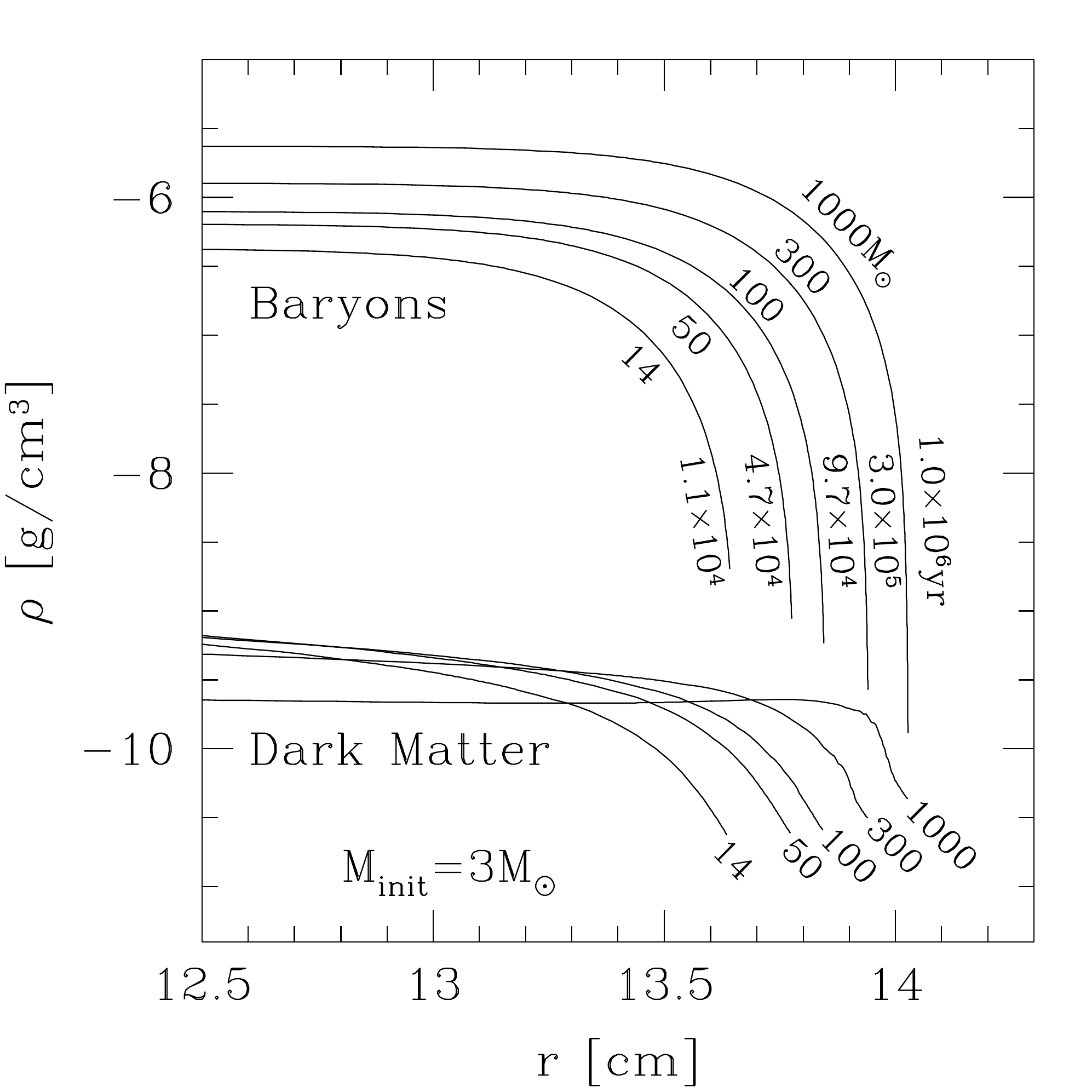}
\caption{Evolution of a dark star (n=1.5) as mass is accreted onto the
  initial protostellar core of 3 M$_\odot$; this figure assumes a constant accretion rate
  $\dot M = 3 \times 10^{-3} M_\odot$/yr.  The set of upper (lower)
  curves correspond to the baryonic (DM) density profile at different
  masses and times. Note that DM constitutes $<10^{-3}$ of the mass of
  the DS.}
\end{figure}

\section{Later stages: Capture}

The dark stars will last as long as the DM fuel inside them persists.  The
original DM inside the stars runs out in about a million years.
However, as discussed in the next paragraph, the DM may be replenished
by capture, so that the DS can live indefinitely due to DS
annihilation.  Capture only becomes
 important once the DS is already large (hundreds of solar masses), and only with the
 additional particle physics ingredient of a significant WIMP/nucleon elastic scattering
cross section at or near the current experimental bounds.

The new source of DM in the first stars is capture of DM particles
from the ambient medium.  Any DM particle that passes through the
DS has some probability of interacting with a nucleus in the star
and being captured. The new particle physics ingredient required
here is a significant scattering cross section between the WIMPs
and nuclei. Whereas the annihilation cross section is
fixed by the relic density, the scattering cross section is a
somewhat free parameter, set only by bounds from direct detection 
experiments.  
Two simultaneous papers  \cite{Freese:2008ur, Iocco:2008xb} found the
same basic idea: the DM luminosity from captured WIMPs can be larger
than fusion for the DS. Two uncertainties exist here: the scattering
cross section, and the amount of DM in the ambient medium to capture
from.  DS studies following the original papers that include
capture have assumed (i)  the maximal scattering cross sections allowed by
experimental bounds and (ii) ambient DM densities that are never depleted.
With these assumptions, DS evolution models with DM heating after the
onset of fusion have now been studied in several papers
\cite{Iocco:2008rb, Taoso:2008kw, Yoon:2008km}.
The two original papers on capture in Pop III stars, as well as these additional papers,
would all apply to later generations of stars (vs. the very first ones), as the stellar masses
on the Zero Age Main Sequence (ZAMS) were taken to be $\sim 100 M_\odot$.

We suspect that the DS will eventually leave their high
density homes in the centers of DM haloes, especially once mergers of
haloes with other objects takes place, and then the DM fuel will run
out. The star will eventually be powered by fusion. Whenever it again
encounters a high DM density region, the DS can capture more DM and be
born again.

In our work, we have considered two separate cases: 1) 'no capture': the case where ambient
density and/or scattering cross section are simply not high enough for capture to matter
and (ii) 'minimal capture": the case
where the stellar luminosity (on the ZAMS) has equal contributions
from DM heating and from fusion.

If the capture rate were much higher, say two or more orders of
magnitude higher than the minimal value considered here,  the star could
 stay DM powered and sufficiently
 cool such that baryons can in principle continue to accrete onto the star
 indefinitely, or at least until the star is disrupted. This latter case will be explored in
  a future paper where it will be shown that the dark star could 
easily end up with a  mass on  the order of several tens of thousands of solar masses 
  and a lifetime of least tens of millions of years.

\section{Results and Predictions}

While DM powers the dark stars, they are cool (surface temperatures less than 10,000K)
and bright ($10^6 L_\odot$).  These properties are very different from standard
Pop III stars, which are $\sim 140 M_\odot$ \cite{mckee} and have surface temperatures
exceeding 30,000K.  One can thus
hope to find DS and differentiate them from standard Pop III stars, e.g. in JWST.

Once the DM fuel runs out inside the DS, the star contracts
until it reaches $10^8$K and fusion sets in.
Our final result \cite{Freese:2008wh} in all cases is very large first
  stars; e.g., for 100 GeV WIMPs, the first stars have $M_{DS} = 800
M_\odot$.  
The implication is that main-sequence stars of Pop. III are very massive.
Regardless of  uncertain parameters such as the DM particle
mass, the accretion rate, and scattering, DS  are cool, massive,
 puffy and extended. The final masses lie in the range 500--1000 M$_\odot$,
very weakly dependent on particle masses, which were taken to vary
over a factor of $10^4$.

One may ask how long the dark stars live.  If there is no capture, they live until the
 DM they are able to pull in via adiabatic contraction runs out; the numerical
results show lifetimes in the range $3 \times 10^5$ to $5 \times 10^5$ yr.
If there is capture, they can continue to exist as long as they
 reside in a medium with a high enough density of dark matter to provide their
entire energy by scattering, capture, and annihilation.

Once the stars are on the Main Sequence, powered by fusion, they will not
last very long before collapsing to form black holes. 
DS would make plausible precursors of the $10^9 M_\odot$
black holes observed at $z=6$ \cite{Li, Pelopessy}; of
Intermediate Mass Black Holes; of black holes at the centers of galaxies; and
of the black holes recently inferred as an explanation of the extragalactic radio excess seen by  the ARCADE experiment \cite{Seiffert}.  However, see
Alvarez et al. (2008) who present  caveats regarding the growth of early black holes.
The final fate of our stars once they reach the MS is uncertain; it is
possible that they could become supernovae \cite{Ohkubo}, leaving
behind perhaps half their mass as black holes. In this case the 
presumed very bright supernova could possibly be observable, and
the resultant black holes could still be important.
In addition, the black hole remnants from DS could play a role in
  high-redshift gamma ray bursts thought to take
place due to accretion onto early black holes (we thank G. Kanbach for 
making us aware of this possibility).

Standard Pop III stars are thought to be $\sim (100-200) M_\odot$, whereas
DS lead to far more massive MS stars.
Heger \& Woosley \cite{HW}    showed that for $140
M_\odot < M < 260 M_\odot$, pair instability supernovae lead to
odd-even effects in the nuclei produced; such element abundances have not been observed.
Other constraints on DS will arise from cosmological considerations.
A first study of their effects (and those of the resultant MS stars) on reionization have been done by Schleicher et al \cite{Schleicher}, and further work in this direction is warranted.

\section{Conclusion}
95\% of the mass in galaxies and clusters of galaxies is in the form
of an unknown type of dark matter.  One of the key properties of WIMP candidates is its
annihilation cross section, yielding the proper relic density
today. As a consequence of this annihilation, the first stars in the
universe may provide another avenue to test the DM hypothesis. These
stars may be powered by DM annihilation, and one can look for them in
upcoming telescopes.  It is an exciting prospect to discover a new type of
star powered by the dark matter in the universe.

In short, the first stars to form in the universe may be Dark Stars
powered by DM heating rather than by fusion.  Our work indicates that
they may be very large ($800 M_\odot$ for 100 GeV mass WIMPs). Once DS
are found, one can use them as a tool to study the properties of WIMPs.

We also briefly mention a separate work \cite{BH}, in which we studied a different possible dark matter
candidate: we studied the effect of primordial black holes on the first stars.  We found that
these small black holes, again adiabatically contracted into the first stars, fall to the center of the
star by dynamical friction.  There they form a single large black hole which can eat the entire star
and accrete from the surrounding medium.  Again we have a mechanism for forming $> 1000 M_\odot$
black holes at early times, which may explain or serve as seeds for the intermediate mass or large black holes found in many places in the universe. 

\section{Acknowledgments} K. Freese thanks her collaborators in this research: Anthony Aguirre, Peter Bodenheimer, Paolo Gondolo, and Doug Spolyar. She also thanks Naoki Yoshida for Figure 1. She ackhowledges support from the DOE and MCTP via the University of Michigan.

\end{document}